\documentclass[aps,prl,twocolumn,superscriptaddress]{revtex4}

\usepackage{graphicx}
\usepackage{amssymb}
\usepackage{enumerate}

\input epsf

\begin{document}

\title{Bragg spectroscopy of a strongly interacting $^{85}$Rb Bose-Einstein condensate}
\author{S. B. Papp}
\author{J. M. Pino}
\author{R. J. Wild}
\author{S. Ronen}
\address{JILA, Quantum Physics Division, National Institute of Standards and Technology
and Department of Physics, University of Colorado, Boulder, Colorado
80309-0440, USA}
\author{C. E. Wieman}
\address{University of British Colombia, Vancouver, BC V6T 1Z1,
Canada}
\address{JILA, Quantum Physics Division, National Institute of Standards and Technology
and Department of Physics, University of Colorado, Boulder, Colorado
80309-0440, USA}
\author{D. S. Jin}
\author{E. A. Cornell}
\email[Email: ]{cornell@jila.colorado.edu}
\address{JILA, Quantum Physics Division, National Institute of Standards and Technology
and Department of Physics, University of Colorado, Boulder, Colorado
80309-0440, USA}
\date{\today}

\begin{abstract}

We report on measurements of the excitation spectrum of a strongly
interacting  Bose-Einstein condensate (BEC).  A magnetic-field
Feshbach resonance is used to tune atom-atom interactions in the
condensate and to reach a regime where quantum depletion and
beyond mean-field corrections to the condensate chemical
potential are significant. We use two-photon Bragg spectroscopy
to probe the condensate excitation spectrum; our results
demonstrate the onset of beyond mean-field effects in a gaseous
BEC.

\end{abstract}
 \maketitle\

The concept of an interacting but dilute Bose gas was originally
developed fifty years ago as a theoretically tractable surrogate
for superfluid liquid helium. For some
years after the eventual experimental realization of dilute-gas
BEC, experiments were performed mainly in the extreme dilute
limit, in which atom-atom correlations were of negligible
significance. Such correlations again assumed a central role,
however, in the 2002 experiments on a Mott state for bosons in an
optical lattice \cite{Greiner2002a} and on atom-molecule
coherence near a Feshbach resonance \cite{Donley2002a}. Atom-atom
correlations are also central to the current hot-topic field of
resonant fermionic condensates \cite{Regal2004a,Zwierlein2004a}.

In this paper we describe an experimental study of elementary
excitations in a system which harkens back to a previous century, in
that, like liquid helium, it is a strongly interacting, bulk,
bosonic superfluid.  Unlike liquid helium, our gas of Bose-condensed
$^{85}$Rb has the modern virtue of Feshbach-tunable interactions
well-described by a scattering length $a$ that is much larger than
the reach of the actual interatomic potential. Our primary tool for
characterizing the sample is Bragg spectroscopy
\cite{Stenger1999b,Stamper-Kurn1999a}.

The convenience of a Feshbach-tunable scattering length in a
bosonic system comes at a cost: at high values of $na^3$ the gas
decays relatively quickly.  For this reason, we are not able to
use, for example, shifts in breathing-mode frequencies
\cite{Pitaevskii1998a} that have been so fruitful in the resonant
fermion systems \cite{Altmeyer2007a, Kinast2004b}. These
measurements require that the gas survive at least as long as it
takes to traverse the sample at the speed of sound.  Bragg
scattering, on the other hand, is essentially a local
measurement. The magnetic field can be ramped to tune the sample
to a large scattering length, and the Bragg pulse can be applied
on a time scale faster than that required for global equilibrium
but still consistent with local many-body quasi-equilibrium.

What should we expect for the energy, $\hbar \omega(k)$, of an
excitation with momentum $\hbar k$ above a Bose-Einstein condensate
of homogeneous density $n$? In the absence of any interactions, the
energy of the excitation is simply $\hbar^2 k^2/(2m)$, where $m$ is
the mass of an atom.  In the simplest description of interactions,
atoms in the condensate feel a chemical potential, $\mu = 4 \pi
\hbar^2 n a/m$ \cite{Dalfovo1999a}. On the other hand, a high
momentum excitation has equal direct and exchange interactions, of
magnitude $\mu$ each, with the low momentum atoms. Thus, the energy
required to promote an atom out of the condensate, less the bare
kinetic energy, is
\begin{equation}
 \hbar \omega (k)- \frac{\hbar^2 k^2}{2m} = 2\mu - \mu= \frac{4 \pi \hbar^2 na}{m}.
\end{equation}
The simple model leading to this energy shift, linear in $n$ and
$a$ (see Fig. 1), relies implicitly on the smallness of three
dimensionless parameters, which are the following:
\begin{enumerate}[1.]
\item
$\sqrt{8\pi na^3} \ll 1$, so that interactions are amenable to a
mean-field treatment.

\item
$ka \ll 1$, so that the two-body scattering amplitude is momentum
independent.

\item
$1/(k\xi) \ll 1$, where $\xi = \hbar/(2m\mu)^{1/2}$ is the healing
length. This limit corresponds to the excitation being cleanly in
the free-particle regime. For larger $1/(k\xi)$,  the excitation
becomes more of a collective object until, in the opposite limit
$1/(k\xi) \gg 1$, the excitation is a phonon.  The crossover regime
was studied experimentally \cite{Steinhauer2002a,Ozeri2005a} with
the results confirming the original theory by Bogoliubov
\cite{Bogolubov1947a}.
\end{enumerate}

The primary goal of the work reported in this paper is to push into
a regime where interactions in a cold Bose gas $\it{can't}$ be
treated as a mean field.  For our data the next order, Lee Huang
Yang (LHY) correction $\alpha \equiv \frac{32}{3\pi\sqrt8} \sqrt{8
\pi na^3}$ \cite{Lee1957a,Lee1957b} is as large as 0.5. As we
increase $a$, the parameters 2 and 3 will also increase. For the
data presented in this paper $1/(k\xi)<0.5$ and $ka$ can be as large
as 0.8. We are not aware of any theoretical treatment that
simultaneously allows all three parameters to be nonzero.
Understanding the regime in which all three parameters are
approximately unity is particularly significant because it
corresponds to the intriguing roton minimum in the dispersion
relation in superfluid liquid He \cite{Nozieres1990a,Griffin1993a}.
The behavior in our gas is in some sense more universal than that in
the liquid, because the scattering length is much larger than the
range of the interatomic potential \cite{Braaten2002a}. We hope that
the results presented below, and follow-up work in progress for
lower values of $k$, will stimulate interest in the problem.

\begin{figure}[htbp]
\begin{center}
\includegraphics[width=80mm]{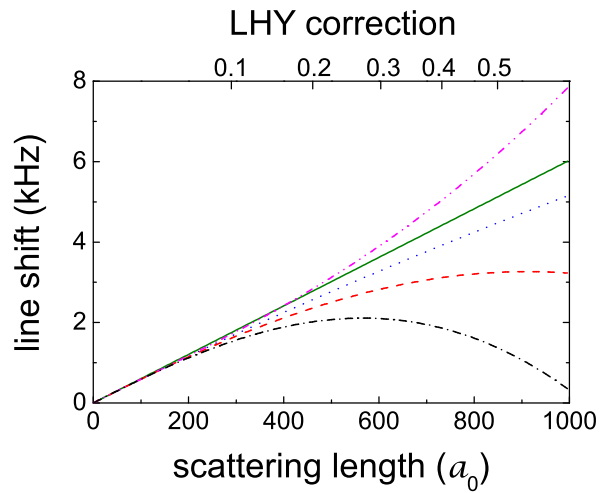}
\caption{
   Theoretical predictions for the peak of the
Bragg resonance less the bare kinetic energy for an excitation of
wave vector $k = 2\,\frac{2\,\pi}{780\, nm}$, in the case of
homogenous density $7.6 \times 10^{13} $ cm$^{-3}$. The theory lines
are described in more detail in the text and correspond to assuming
the following quantities are small: $ka$, $\sqrt{8\pi na^3}$ and
$1/(k\xi)$ (green, solid); $ka$ and $\sqrt{8\pi na^3}$ (blue, dots);
$ka$ (magenta, dashdotdot);  $\sqrt{8\pi na^3}$ (red, dashes); and
$1/(k\xi)$ (black, dashdot). For reference the top axis shows the
size of the LHY correction, $\alpha$, relative to the condensate
chemical potential. \label{theory}}
\end{center}
\end{figure}

 In Fig. \ref{theory}, we plot various calculations good to
lowest order in combinations of at most two of the three small
parameters.  Beliaev gives a result for $ka$ and $1/(k\xi)$ being
nonzero (Eqn. 4.7 in Ref. \cite{Beliaev1958a}), and numerical
integration of Eqns. 5.13-5.14 in Ref. \cite{Beliaev1958a} (see
also Ref. \cite{Mohling1960a}) yields a result for $1/(k\xi)$ and
$\sqrt{8 \pi na^3}$ being nonzero; both results are plotted in
Fig. 1.   We plot also in Fig. 1 the result of a simple model
accounting for finite $ka$ and $\sqrt{8\pi na^3}$ in the limit of
vanishing $1/(k\xi)$. Here, Eqn. 1 is modified to become
\begin{equation}
   \hbar \omega (k) - \frac{\hbar^2 k^2}{2m} = \frac{8\pi \hbar^2 na(k/2)}{m}-(1+\alpha) \frac{4 \pi
\hbar^2 n a}{m}.
\end{equation}
The argument is simply that $1/(k\xi)\to0$ means the excitation $k$
is very large compared to the momentum content of the correlations
induced by nonzero $\sqrt{8\pi na^3}$. Thus, for the interaction
energy felt by the excitation we ignore many-body corrections and
simply use $8\pi \hbar^2 na(k/2)/m$, where $a(k/2) = - Re (f(k/2))$
and $f(k/2)$ is the two-body symmetrized scattering amplitude for
relative wave vector $k/2$. For the energy of an atom in the
condensate, we replace $\mu$ with $(1+\alpha)4 \pi \hbar^2 n a/m$.

We don't expect Eqn. (2) to be valid in the regime of our
experiment, because our value of $k$ is insufficiently large to
ensure the excitations are unaffected by the correlations induced in
the condensate by $\sqrt{8\pi na^3}$ effects.  Numerically
evaluating the poles of the Green's function, Eqns. 5.13-5.14 in
Ref. \cite{Beliaev1958a}, shows that our Eqn. (2) applies only when
$1/(k\xi) \lesssim 0.05$, whereas we reach $1/(k\,\xi)=0.5$. At high
momentum (small $1/(k\xi)$), the effect of quantum depletion is to
shift the dispersion curve downwards relative to the condensate
energy, while at very small momentum, the LHY enhancement to $\mu$
increases the speed of sound and thus increases low-$k$ values of
$\omega(k)$. The up-shift due to $\sqrt{8\pi na^3}$ persists even to
the intermediate momenta of our experiment, which is why the
dashdotdot curve is above the solid line in Fig. 1.

Our experiments are performed using a $^{85}$Rb BEC near a Feshbach
resonance at 155 G \cite{Cornish2000a,Roberts2000a}.   A gas of
$^{85}$Rb atoms in the $|F=2, m_F=-2\rangle$ state is first
sympathetically cooled with $^{87}$Rb in a magnetic trap and then
evaporated directly to ultralow temperature in an optical dipole trap
\cite{Papp2006a}.  We create a single-species $^{85}$Rb condensate
\cite{note}, with $40,000$ atoms and a condensate fraction greater than
85\%, in a weakly confining optical dipole trap at a magnetic field
above the Feshbach resonance.  Curvature of the magnetic field
enhances confinement along the axial direction of the optical trap.
Following evaporative cooling, the optical dipole trap is
recompressed and the final trap has a measured radial (axial) trap
frequency of $2\pi\times134$ Hz ($2\pi\times2.9$ Hz), yielding a
condensate mean density of $2.1\times10^{13}$ cm$^{-3}$.

Bragg spectroscopy via stimulated two-photon transitions provides a
direct probe of the condensate excitation spectrum.  Two
counter-propagating, near-resonant laser beams are aligned along the
long axis of the condensate.  The momentum imparted to a condensate
excitation is given by $\hbar k = 2\,\hbar\,k_L $ where $k_L =
\frac{2\,\pi}{780\, nm}$ is the wave vector of a beam. The
excitation energy is scanned by adjusting the frequency difference
of the two laser beams. The average of the two frequencies is red
detuned from atomic resonance by 4.2 GHz.  The intensity and pulse
duration of the Bragg beams are chosen so that the fraction of the
condensate excited is less than 10\%.

Just before performing the Bragg spectroscopy, we transiently
enhance the condensate density by means of large amplitude radial
and axial breathing modes, which we excite by modulating the
magnetic field and thus the Feshbach-modified scattering length. The
rates of the ramps are limited so the $\dot{a}/a$ never exceeds $0.06
\hbar /(m a^2)$.  The scattering length is derived from measurements
of the magnetic field and a previous measurement of the $^{85}$Rb
Feshbach resonance \cite{Claussen2003a}. Sychronized with the inner
turning point of the radial oscillation, we ramp the scattering length
to the value for a given measurement and then pulse on the Bragg
beams. During the pulse, the cloud's inward motion
is checked and it begins to breathe outward. We model the resulting
time-dependent condensate density  using a variational solution to
the Gross-Pitaevskii equation \cite{Perez-Garcia1997a}, which predicts
that the density of the cloud does not change by more than
$30\%$ during the Bragg pulse. We can meet this goal only by using
progressively shorter Bragg pulses for higher values of desired $a$.
The time- and space-averaged density during the pulse is
approximately $7.6\times10^{13}$ cm$^{-3}$, but this depends weakly
on the final value of $a$.

After the Bragg pulse, we ramp $a$ to a large, fixed value to ensure
that the momentum of the excitations is spread via collisions
\cite{Katz2002a,Chikkatur2000a} to the entire condensate sample.  We
then infer the total momentum, and thus excitation fraction, from
the amplitude of the resulting axial slosh, measured via an absorption
image taken of the cloud at a time near its axial turning point.

\begin{figure}[htbp]
\begin{center}
\includegraphics[width=90mm]{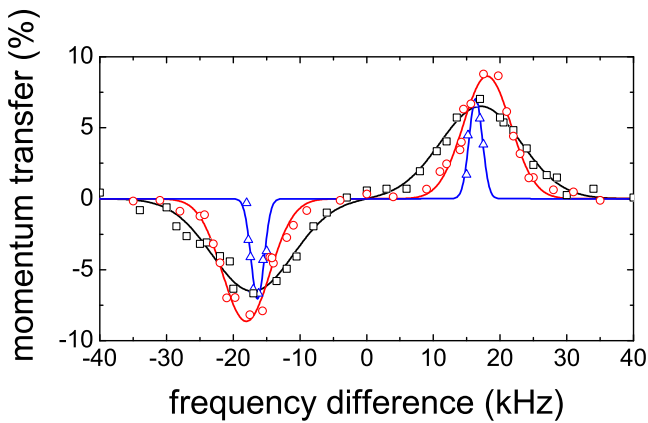}
 \caption{Typical Bragg spectra at
       a scattering length of 100 $a_0$ (blue triangles), 585 $a_0$
       (red circles), and 890 $a_0$ (black squares).  The excitation
       fraction is determined from the measured momentum transferred
       to the BEC and plotted as a function of the frequency
       difference between the two Bragg beams.
       Lines are fits of the data as described in the text. Mean-field theory
predicts a continuous increase in the line shift with increasing
$a$, however by 890 $a_0$ our data display a \textit{decreasing}
shift with stronger interactions. \label{example}}
\end{center}
\end{figure}

\begin{figure}[htbp]
\begin{center}
\includegraphics[width=90mm]{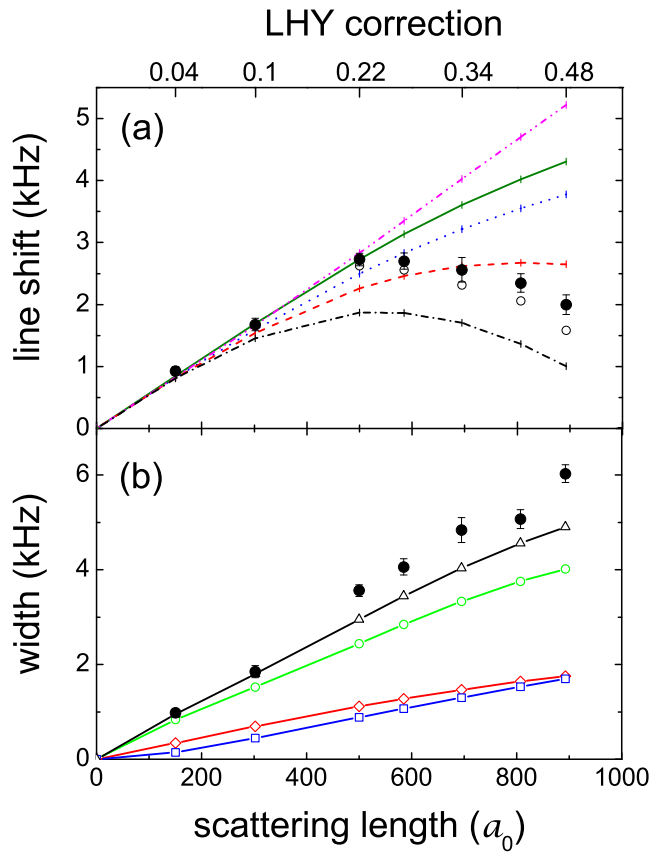}
\caption{(a) Bragg line shift and (b) width as a function of
scattering length.  The hollow circles are our observations. The solid circles are data
corrected for a fitting systematic associated with the broad thermal atom
background, and the error bars represent fit uncertainties. The theory
lines and LHY correction in (a) are as in
Fig. 1 except they are calculated for the trapped gas using a local
density approximation for each of the corresponding data points; the
mean BEC density ranges from $6.3\times10^{13}$ cm$^{-3}$ to
$7.6\times10^{13}$ cm$^{-3}$. In (b) the solid black circles are the
rms width of a gaussian fit to the Bragg spectra. Black triangles
are from a fit to a convolution of various contributions to the
width calculated under the conditions
   of our measurements.  The remaining symbols characterize constituent
    contributions to the convolution including the Lorentzian
    FWHM width due to collisions (blue squares), and
     the RMS widths of contributions due to the inhomogeneous density (red diamonds)
      and the pulse duration (green circles).  The largest contribution to
      the width comes from the pulse duration; because the jump to large $a$
       initiates rapid expansion of the BEC, ever shorter pulses are
      used to obtain the spectra at larger $a$.
 \label{expt}}
\end{center}
\end{figure}

 Figure \ref{example} shows measured Bragg
spectra for three values of $a$. We fit each Bragg spectrum to an
antisymmetric function assuming a Gaussian peak and extract a center
frequency and an RMS width. The Bragg line shift is the difference
between the fitted center and the ideal gas result
$\frac{1}{2\pi}\frac{\hbar\,k^2}{2\,m}=15.423$ kHz. In Fig.
\ref{expt}a we plot our measured line shifts as a function of the
scattering length $a$. For $a\lesssim300 a_0$ (where the predicted
LHY correction is already a 10$\%$ effect), the measured line shift
($\bullet$ in Fig. \ref{expt}a) agrees with the simple mean-field
result (Eqn. 1). However, as the scattering length is increased
further, the resonance line shift deviates significantly from the
mean-field prediction. The measured line shift reaches a maximum
near $a=500 a_0$ and then \textit{decreases} as the scattering
length is increased further.

At large $a$ we find that our measured line shift exhibits a
systematic dependence on the temperature of the sample
\cite{Brunello2001a}. Non-condensed $^{85}$Rb atoms also respond to the
Bragg pulse, and this causes an observable effect in the measured
line shift when the spectral width of the condensate response
becomes comparable to that of the non-condensed atoms (for $a > 500
a_0$). We vary the temperature of the gas to characterize this
effect and we apply a small correction to our data to represent the
expected line shift at zero temperature ($\bullet$ in Fig.
\ref{expt}a).

The theory curves of Fig. 1 are adapted for direct comparison with
our data (Fig 3a) by calculating, at each tick mark on the curves, a
spatial and temporal average over a local density approximation for
the corresponding data point, assuming the density of the condensate
is given by a time-varying Thomas-Fermi profile. In these
predictions, we account for observed number losses (typically $< 30\%$)
during our experiments.  The observed loss is consistent with the
previously observed $a$-dependent three-body recombination rate of
$^{85}$Rb \cite{Roberts2000a}.

Figure \ref{expt}b shows the measured width of the Bragg peak as a
function of $a$.  Several effects contribute to the total width of
the Bragg resonance including the finite duration of the pulse, the
inhomogeneous density of the trapped condensate \cite{Stenger1999b},
collisions between the excitations and the condensate
\cite{Katz2002a}, and Doppler broadening \cite{Stenger1999b}. In our
case, Doppler broadening is negligible since the axial size of the
condensate is relatively large.  To understand the total width we
convolve the various calculated lineshapes of the remaining three
effects.  We fit a Gaussian to the convolution (to match the
Gaussian fit to our data) and the RMS width from this fit is shown
in Fig. \ref{expt}b (black $\triangle$). Fig. \ref{expt}b also shows
the expected contributions to the width from each of the three
effects. For the lineshape due to collisions we expect a Lorentzian
with a full width at half maximum $\delta\nu =
\frac{1}{2\,\pi}\frac{n\,\sigma\,k}{m}$, where $\sigma$ is the
elastic cross section for collisions between the excitations and low
momentum atoms. In calculating $\sigma$ we include
the suppression in the phonon regime predicted by
Beliaev \cite{Beliaev1958a,Katz2002a}.  The measured Bragg width
exceeds the predicted width in the strongly interacting regime.
However, many of the theoretical difficulties in describing the line
shift apply also to predicting the width that arises from inhomogeneous
density and excitation lifetime.

In comparing our measured line shifts with the various theory models
(Fig. 3), one should remember that none of the models takes into
account the fact that the inter-atomic potential supports bound
states, whose existence means that a Bose-condensed gas can be only
a quasi-equilibrium state for any atomic species.  This is
especially relevant for increasing values of $a$. A key future goal
of our work is to experimentally explore different time-scales for
the establishment of local many-body quasi-equilibrium, and for
longer-term evolution. The extreme aspect ratio of our sample
hastens the loss of density that occurs during the expansion caused
by the ramp to high $a$. At present, we are required to use short
duration Bragg pulses, which limits our spectral resolution, and we are
prevented from tracking the time evolution of line shifts. An ongoing
redesign to a more spherical geometry will help. In addition, the
Bragg beams are being reconfigured to allow access to the low-$k$, pure-phonon
regime for which $1/(k\,\xi) \gg 1$.

We gratefully acknowledge useful conversations with J. Bohn, M. Holland, R.
Ballagh, S. Stringari and the JILA ultracold atom collaboration.
This work is supported by NSF and ONR.


\begin{thebibliography}{10}
\bibitem{note}
The number of trapped $^{87}$Rb atoms remaining is $< 10\%$ of the $^{85}$Rb number.
\end{thebibliography}
\end{document}